\documentclass{elsart}

\newcommand{\be}{\begin{equation}}
\newcommand{\ee}{\end{equation}}
\def\ap{\approx}
\def\ba{\begin{eqnarray}}
\def\ba{\begin{eqnarray}}
\newcommand{\lsim}   {\mathrel{\mathop{\kern 0pt \rlap
  {\raise.2ex\hbox{$<$}}}
  \lower.9ex\hbox{\kern-.190em $\sim$}}}
\newcommand{\gsim}   {\mathrel{\mathop{\kern 0pt \rlap
  {\raise.2ex\hbox{$>$}}}
  \lower.9ex\hbox{\kern-.190em $\sim$}}}

\begin{document}
\begin{frontmatter}

\title{Ultra-high energy LSP}

\author{V. Berezinsky$^1$ and M. Kachelrie{\ss}$^{1,2}$}

\address{$^1$INFN, Laboratori Nazionali del Gran Sasso,
             I--67010 Assergi (AQ), Italy}
\address{$^2$Theoretische Physik I, Ruhr-Universit\"at Bochum, 
             D--44780  Bochum, Germany}

\begin{abstract}
We argue that the lightest supersymmetric particles (LSP) can be produced with
extremely high energies $E\gsim 10^{10}$~GeV in the Universe at the
present epoch. Their most probable sources are decaying superheavy particles 
produced by
topological defects or as relic Big Bang particle. We
discuss the mechanisms of production of LSP at ultra-high energies (UHE)
and the interaction of the  UHE LSP with matter. The most 
attention is given to the neutralino as LSP, although the gluino is
also considered as a phenomenological possibility.
\end{abstract}
\begin{keyword}
Supersymmetry, cosmic rays, topological defects.
\end{keyword}
\end{frontmatter}

\section{Introduction}
Cold Dark Matter (CDM) is probably the most abundant form of matter in 
the Universe. A natural CDM candidate is the lightest
supersymmetric particle (LSP) which is stable, if $R$-parity is conserved. 
In this {\em Letter}, we address the question if
the LSP can be also among the ultra-high energy (UHE) particles
filling the Universe. 
Theoretically the best motivated candidates for LSP are the neutralino and 
gravitino. We shall not consider the latter, because it is practically 
undetectable  as UHE particle. Therefore, we pay most attention to the
neutralino. 

In all elaborated SUSY models the gluino is not the LSP. Only, if
the dimension-three SUSY breaking terms are set
to zero by hand, gluino with mass $m_{\tilde g}={\mathcal
O}(1~{\rm GeV})$ can be the LSP \cite{fa96}. There is some controversy if
the low-mass window $1~{\rm GeV} \lsim m_{\tilde g} \lsim 4$~GeV for
the gluino is still allowed \cite{pdg,aleph}. Nevertheless, we shall study
the production of high-energy gluinos and their interaction with matter
being inspired by the recent suggestion \cite{ch97} (see also \cite{mo97}), 
that the atmospheric showers observed at the highest energies can be
produced by colourless hadrons containing gluinos. 
We shall refer to any of such hadron as
$\tilde{g}$-hadron ($\tilde{G}$). Light gluinos as 
UHE particles with energy $E \gsim 10^{16}$~eV were  considered in 
some detail in the literature in connection with Cyg X-3 \cite{aur,BI}.
Additionally, we consider heavy gluinos  with $m_{\tilde g} \gsim
150$~GeV \cite{mo97}.

Three mechanisms for the production of UHE LSP can be identified.\\
(i) They can be produced in "astrophysical accelerators" due to the
interaction of accelerated protons with ambient gas. This mechanism 
effectively works 
only in the case of a light gluino, and it was exploited in 80s during the 
Cyg X-3 epic where the {\em glueballino\/} ($\tilde{g}g$ bound state)
was one of the main characters \cite{aur,BI}.
For UHE $\tilde{g}$-hadrons this production mechanism was indicated in Ref.
\cite{ch97}.
The main difficulty of this mechanism is the low LSP flux, which is caused 
by the small cross-section for gluino production 
if $m_{\tilde g} > 1$~GeV and by the low density of the target 
nucleons or photons around the "accelerator". \\
(ii) Evaporating black holes are another possibility.  High energy 
particles can be produced during the final stages of evaporation;  the 
resulting spectra of cosmic rays have been discussed in Refs. \cite{turn}.  
However, when these spectra are combined with various observational 
bounds on the mass fraction of the universe in black holes (see \cite{nov}
for 
a review), one finds that the UHE CR flux from black holes is well below 
the observed flux.\\ 
(iii) Decays of the supermassive particles produced either by topological 
defects or by Big Bang as relic particles, can naturally provide the large 
fluxes of UHE LSP.  

The plan of this paper is as follows: We first discuss the possible sources
of UHE LSP. Then we examine the cascade-production of LSP and their
resulting spectrum. 
After that, we calculate the LSP fluxes for the two most
promising sources. Finally, we examine
the interactions of the UHE LSP with matter and discuss the status of
(quasi-)stable gluino.

\section{Topological defects and supermassive relic particles}
Topological defects and
supermassive relic particles are the two most promising sources.  
Topological defects \cite{td} such as superconducting
strings,  monopoles, and monopoles connected by strings can produce 
UHE particles \cite{CR_TD}. 
Here we will concentrate on {\em cosmic necklaces\/}, since this model 
seems to provide the largest UHE particle flux for fixed density of 
electromagnetic cascade radiation. 

Cosmic necklaces are hybrid defects
consisting of monopoles connected by a string. These defects are produced 
by the symmetry breaking $G\to H\times U(1) \to H\times Z_2$.
In the first phase transition at scale $\eta_m$, monopoles are 
produced. At the second phase transition, at scale  $\eta_s<\eta_m$, each 
monopole gets attached to two strings. The basic parameter for the evolution
of necklaces is the ratio of monopole mass and the mass of the string 
between two monopoles, $\mu d$, where $\mu \sim \eta_s^2$ is the mass 
density of the string and $d$ the distance between two monopoles.
Strings loose their energy and can contract due to gravitational radiation.
As a result, all monopoles annihilate in the end producing
superheavy Higgs, gauge bosons and their supersymmetric partners 
which we call collectively $X$-particles.
The rate of $X$-particle production can be estimated as
\be
 \frac{\d n_X}{\d t} \sim \frac{r^2 \mu}{t^3 m_X} \,.
\label{n/t}
\ee
Similar to the case of UHE protons, the flux of UHE LSP is determined
mainly by two 
parameters, $r^2\mu$ and $m_X$, which values must be of order 
$10^{27}$~GeV$^2$ and $10^{14}$~GeV, respectively, to have 
the flux close to the observed one.
For a more complete discussion see Ref.~\cite{BV}.   
The diffuse flux of LSP produced by the decay of $X$-particles from necklaces
is given by
\be
I_{\rm LSP}(E)=\frac{1}{4\pi} \, R(E) \, \frac{\d n_X}{\d t}  
                                \frac{\d N_{\rm LSP}(E)}{\d E} \,,
\label{flux}
\ee
where $\d N_{LSP}/\d E$ is the spectrum of LSP from the decay of
$X$-particles. 
Furthermore, $R(E)$ is the attenuation length $\lambda(E)$ of the LSP if
$\lambda(E)<ct_0$ and $R(E)=ct_0$ otherwise, where $t_0$ is the age of the 
Universe.

If the neutralino $\chi$ is the LSP, $\lambda(E)$ is much larger 
than $ct_0$. It is determined as 
$\lambda(E)=(1/E \cdot \d E_{\rm loss}/\d t)^{-1}$ 
and the dominant contribution is given by the   
scattering of neutralino off background neutrinos, 
$\chi+\nu_{\rm BB}\to\chi+\nu$. The largest cross-section,
$\sigma \sim G_F^2(E/m_{\chi})^2\epsilon^2_{\nu}$, where $\epsilon_{\nu}$ 
is the energy of the relic neutrino, is provided by the Higgsino component 
of the neutralino with $Z^0$-exchange in the $t$-channel. 
Apart from the smallness of the cross-section, $\lambda$ is further
reduced by the small energy transfer in one collision. The Universe 
becomes transparent for UHE neutralino at red shift $z<1\cdot 10^4$, 
however production of neutralinos at early cosmological epochs is not 
important in the case of topological defects.

The dominant energy-loss process of the $\tilde{g}$-hadron is pion 
production in collisions with microwave photons. Pion production
effectively starts 
at the same Lorentz-factor as in the case of the proton. This
implies that the energy of the GZK cutoff is a factor $m_{\tilde{g}}/m_p$
higher than in case of the proton. The attenuation length also 
increases because the fraction of energy lost near the threshold of 
production is small, $\mu/m_{\tilde{g}}$, where $\mu$ is a pion mass.
Therefore, even for light $\tilde{g}$-hadrons, $m_{\tilde{g}} \gsim
2$~GeV, the steepening of the spectrum is less pronounced than for protons.

Let us now come to superheavy relic particles as a source of UHE LSP.
This case is formally identical to the production of $X$-particles 
by topological defects if one replaces the RHS of Eq.~(\ref{n/t})
by $n_X /\tau_X$.
Following Ref.~\cite{BKV} we will use the ratio $r_X = \xi_X t_0/\tau_X$, 
$\xi_X =\rho_X/\rho_{\rm CDM}$ as parameter characterizing the model (here
$\rho_X $ and $\rho_{\rm CDM}$ are the mass density of $X$-particles
and of the total CDM, respectively).  
The flux can be calculated by Eq.~(\ref{flux}), where now $R(E)$ is 
the size of galactic halo $R_h$. We adopt the 
following astrophysical parameters for this case \cite{BKV}:
$R_h=100$~kpc, $\Omega_{\rm CDM}=0.2h^2$ and $h=0.6$.
We assume for the halo density $m_Xn_X^h=\xi_X\rho_{\rm CDM}^h$ and
for the extragalactic density
$m_Xn_X^{\rm ex}=\xi_X\Omega_{\rm CDM}\rho_{\rm cr}$,
where $\xi_X$ describes the fraction
of $X$-particles in CDM and $\Omega_{\rm CDM}$ is the CDM
density in units of the critical density $\rho_{\rm cr}$.

\section{Cascade-production of LSPs at the decay of $X$-particle}
The decay of $X$-particle results in a particle cascade similar to the QCD 
cascade in $e^+ e^-$ annihilation. The basis of the cascade development in 
both cases is given by probability $p$ of production of an extra parton,
$p \sim g^2 \ln Q^2 >1$, where $g$ is a coupling constant.
In the process of the cascade development,
the energy and the virtuality $Q^2$ of the
cascade particles diminish  progressively. Until the virtuality $Q^2$
remains larger than the SUSY scale $M_{\rm SUSY}^2 \sim (1~{\rm
TeV})^2$, the decay 
channels to the usual particles and their supersymmetric partners have
equal probabilities and the number of SUSY 
particles at each generation is exactly equal to that of usual particles.
When $Q^2$ reaches $M_{\rm SUSY}^2$, the supersymmetric particles go out of 
equilibrium and decay to the LSP.

We performed a simplified Monte-Carlo simulation\footnote{A more
complete Monte-Carlo simulation of the cascade is in preparation
\cite{casc}.} of the cascade
including as elementary processes the transition probabilities between
fermions, sfermions, gluons, gluinos, $W$-bosons and winos, photons
and neutralinos with probabilities similar to that in QCD. 
For simplicity, we assumed a common mass $M_{\rm SUSY}=1$~TeV for all
SUSY particles except the LSP. We followed  
the evolution until a particle reaches the virtuality 
$Q^2 \leq M_{\rm SUSY}^2$. In the case of the neutralino as LSP, each
SUSY particle is then turned into neutralino after one or several decays.
The case of heavy gluino, $m_{\tilde{g}}\gsim 100$~GeV is very similar, with 
the obvious difference that gluino is turned into $\tilde{g}$-hadron  at the 
confinement radius. 

The case of a very light gluino is different. After SUSY particles go out 
of equilibrium, gluinos still participate in the QCD cascade due to 
$g \to \tilde{g}+\tilde{g}$. The processes of gluino production and 
radiation of gluons by gluinos are very similar to that of quarks and one can 
expect that in the low-energy part of the cascade the number of
gluinos is roughly
equal to the number of quarks. However, in the problem considered here, 
we are not interested in particles with too small $x$ and  
our simulation does not include this low-energy regime.

The obtained spectrum of the LSP can be well approximated for the
energies at interest $E\gg M_{\rm SUSY}$ by
\be 
 \frac{\d N}{\d E} \sim k \,\frac{1}{M_X} \left(\frac{E}{M_X}\right)^{-1.4}
\ee
with $k\sim 0.25$ for both neutralino and gluino as LSP.
The fraction of energy transferred to the LSP is 
$f_{\rm SUSY}\sim 0.4$. In Fig.~1 and 2 we show the resulting LSP fluxes 
from decaying $X$-particles and cosmic necklaces, respectively.
Moreover, the neutrino fluxes calculated in MLLA approximation and
experimental data \cite{data} are shown.

\section{Interaction of UHE neutralino with matter}
Let us consider the interactions of the neutralino $\chi$ 
relevant for their detection. They are somewhat similar to the calculations
\cite{photino} for a photino. Mainly two processes are important
 for its interaction with matter, namely 
the neutralino-nucleon scattering
$\chi+N\to$ all and resonant production of selectron off electrons 
$\chi+e \to \tilde{e} \to$ all.  The first process is based on the
resonant subprocess $\chi+q \to \tilde{q} \to$ all and on neutralino-gluon
scattering. The latter subprocess is important, because for
high energies, and consequently for small scaling variable $x$, the gluon 
content of the nucleon increases fast.

The resonant cross-section of the parton process 
$\tilde\chi + q \to \tilde q_{L,R} \to$ all is 
given by the Breit-Wigner formula as 
\be   
 \sigma^{\rm res} (\hat s )= \frac{\pi \hat s}{\hat p_{\rm cm}^2} \; 
      \frac{\Gamma^2 (\tilde q_{L,R} \to q + \tilde\chi)}
           {(\hat s-M_{L,R}^2)^2 + M_{L,R}^2 \Gamma^2_{\rm tot}} \,,
\label{BW}
\ee 
where $s=2E_\chi m_N$, $E_\chi$ denotes the energy of the incident
neutralino, $m_N$ the nucleon mass,  $M_{L,R}$ are the masses of left-
and right-chiral squarks and $\hat s=sx$.
The total decay width $\Gamma_{\rm tot}$ of the squark  depends
strongly on the mass spectrum of the model.
In the following, we parameterize our ignorance by
$\Gamma_{\rm tot} = z \Gamma (\tilde q_{L,R} \to q + \chi)$ with
$z\geq 1$. 
To obtain the neutralino-nucleon cross-section,  the
parton cross-section $\sigma^{\rm res} (\hat s)$ has to be integrated
with quark distribution functions $q_i(x, \hat s)$ over $x$, down to
$x_{\rm min}= M_{L,R}^2 /s$, and summed over all quarks.
%
%
The total $\chi$-nucleon cross-section due to the neutralino gluon scattering
can be  similarly obtained integrating the parton cross-section  
with the gluon structure function of the  nucleon.

The soft breaking terms of the MSSM are characterized by the
following basic parameters \cite{be96}: 
the masses of $i$ scalar fields, $m_0^i$,
at the GUT scale, the masses of $j$ gaugino fields, $m_{1/2}^j$, also at 
GUT scale,
the ratio of the two vev's of the Higgs fields
$\tan\beta= v_2/v_1$, and the Higgs mixing parameter $\mu$.
To proceed, we choose two different scenarios. In the first one,
we assume universality of scalar masses $m_0$ and fermion masses $m_{1/2}$ at 
the GUT scale. In this case the neutralino is gaugino dominated in
most part of the parameter space of the 
MSSM. To obtain a lower bound for the cross-sections we use the 
configuration with the 
largest values of parameters compatible with a no-fine tuning condition 
\cite{be96}
($m_0 =308\:$GeV, $m_{1/2}=390\:$GeV, and $\mu=561$).
We fix $\tan\beta=8$ throughout.
This corresponds to a neutralino with mass $M_\chi = 160$~GeV and
a gaugino part of $P=Z_{11}^2+Z_{12}^2=0.99$. 
The masses of the squarks are $M_{L,R} \ap 1000$~GeV, except the lightest
stop which has $M\ap 780$~GeV.

In the second scenario, we break universality 
for the two Higgs doublets, keeping the universal value $m_0$ for all other 
scalars and $m_{1/2}$ for gauginos. As a result  
mixed and Higgsino dominated configurations appear \cite{be96}.
We choose the configuration with minimal gaugino part, which is
obtained for $m_0 =1204\:$GeV, $m_{1/2}=295\:$GeV,
$\mu=108$.
The masses of the squarks of the first two generations are 
$M_{L,R} \ap 1400$~GeV, while masses of the third generation range from
$1400-730$~GeV.

The resulting cross-sections are shown in Fig.~3 for 
the case of universality (gaugino dominated neutralino), and in 
Fig.~4 for a non-universal case and a higgsino-dominated 
neutralino. In each figure, the cross-section 
of gluon-neutralino scattering (solid line) and the cross-sections 
of resonant quark-neutralino scattering (dashed lines) for 
$\Gamma_{\rm tot}=z \Gamma(\tilde q_{L,R} \to q + \tilde\chi)$ with
$z=1$ (top) and $z=10$ (bottom) are shown. 
The cross-sections of all subprocesses start to grow at 
energies $s\gg 10^6 \:$(GeV)$^2 \ap M_{L,R}^2$. The rise with $s$ is
caused by the decrease of $x_{\rm min}= M_{L,R}^2/s$ and
$x_{\rm min}= (M_{L,R} + m_q)^2/s$, and the corresponding decrease of
the number of partons with sufficient momentum in the nucleon.
If squarks do not decay mainly into neutralino, 
{\it i.e.\/} $\Gamma_{\rm tot}\gg \Gamma(\tilde q_{L,R} \to q + \tilde\chi)$,
neutralino-gluon scattering gives in both cases the dominant
contribution to the total cross-section. 
At energies $s\ap 10^{10} \:$(GeV)$^2$ or $E_\chi \ap 5 \cdot 10^{18}$eV, 
the neutralino-nucleon cross-section is about $10^{-35}-10^{-34}\:$cm$^2$, 
{\it i.e.\/} slightly lower than the neutrino-nucleon cross-section. 

Let us consider now $\chi +e \to \tilde e\to$~hadrons which is similar to the
Glashow 
resonant scattering $\bar\nu_e +e\to W \to \mu+\bar\nu_\mu$. 
The resonant energy of the neutralino is
\be
E_{\chi}=M_{\tilde e}^2/(2m_e)=9.8\cdot 10^8 (M_{\tilde e}/10^3~{\rm
GeV})^2~{\rm GeV}
\ee  
and the cross-section is also given by Eq. (\ref{BW}).
Now, $\Gamma_{\rm tot}$ is the total decay width of $\tilde{e}$ determined by 
the decay channels to electron and neutralino and neutrino and 
chargino and $M$ is the selectron mass. For the frequency of events
produced by the neutralino flux $I_{\chi}(E)$ in a detector with $N_e$
target electrons one easily obtains 
\be
\nu=\Omega N_e E_0 I_{\chi}(E_0) \sigma_{\rm eff},
\label{freq}
\ee
where $\Omega$ is the solid angle and
\be 
\sigma_{\rm eff}=2\pi \alpha_{\rm ew} \kappa^2 /M_{\tilde e}^2 
                \sim 2\cdot 10^{-35}~{\rm cm}^2
\ee 
is the effective cross-section. The factor $\kappa$ depends on the
composition of the neutralino and is assumed to be of order one.
The resonant events with frequency (\ref{freq}) are produced as a narrow 
peak at $E_{\chi}$ and give an unique signature for neutralino.

\section{Interactions of $\tilde{g}$-hadrons and its status as LSP}
The interaction of UHE $\tilde{g}$-hadrons ($\tilde{G}$) was already
considered in some detail in Ref.~\cite{BI} for the case of glueballino.
Two values determine the interaction of a UHE $\tilde{g}$-hadron with
a nucleon. The first one is the radius of the $\tilde{g}$-hadron. 
This radius is inversely proportional
to the reduced mass of the system and estimates of Ref.~\cite{BI} give for 
the total $\tilde{G}$N-cross-section $\sigma \sim 1$~mb. 
In Ref.~\cite{mo97}, it is argued that this cross-section is of order 
$\Lambda_{QCD}^{-2} \sim 10$~mb. 
However, for production of EAS in the atmosphere only interactions with
large energy transfer are effective. For a gluino with mass 
$m_{\tilde{g}} \sim 3$~GeV or $m_{\tilde{g}} \gsim 150$~GeV the large 
energy transfer corresponds to scattering with large $Q^2$ and thus to small
cross-section. 

Now we consider the diffractive interaction of UHE $\tilde{g}$-hadron with 
nucleons. Then $\tilde{g}$-hadrons exchange with nucleons the
4-momentum $Q$ in the $t$-channel fragmentating into jets of hadrons 
(including $\tilde{g}$-hadron). For a given energy transfer $y=(E-E')/E$,
\be
 Q^2_{\rm min}=M^2y \left( (1-y)^{-1}-m_{\tilde{G}}^2/M^2\right),
\label{Q}
\ee  
where $M > m_{\tilde{G}}$ is the invariant mass of the fragmentation jet.
Unless $y$ is very small, $Q^2_{\rm min}$ is large and the corresponding 
cross-section $\sigma(y) \sim 1/Q^2_{\rm min}(y)$ is small.
We conclude thus that a heavy $\tilde{g}$-hadron ($m_{\tilde{G}}\sim
3$~GeV or $m_{\tilde{G}} \gsim 150$~GeV)
behaves in the atmosphere like a penetrating particle, while a very light 
$\tilde{g}$-hadron interacts like a nucleon.

Let us discuss now the status of the gluino as LSP. 
Accelerator experiments give the lower limit on the gluino mass as 
$m_{\tilde{g}} \gsim 150$~GeV \cite{pdg}. The upper limit of the 
gluino mass is given by cosmological and astrophysical constraints, as 
was recently discussed in \cite{mo97}. In this work it 
was shown that if the gluino provides the dark matter observed in our galaxy,
the signal from gluino annihilation and the abundance of anomalous heavy 
nuclei is too high. Since we are not interested in the case when gluino 
is DM particle, we can use these arguments to obtain an upper limit for 
the gluino mass. Calculating the relic density of 
gluinos (similar as in \cite{mo97}) and using
the condition $\Omega_{\tilde{g}} \ll \Omega_{\rm CDM}$, we obtained 
$m_{\tilde{g}} \ll 9$~TeV.

Now we come to a very strong argument against the existence  
of a light stable or quasistable gluino \cite{VO}.
It is plausible that the glueballino ($\tilde{g}g$) is the lightest hadronic 
state of gluino \cite{aur,BI}. However, {\em gluebarino\/}, i.e. the bound 
state of gluino and three quarks, is almost stable because baryon number 
is extremely weakly violated. In Ref.~\cite{VO} it is argued that the 
lightest gluebarino is the neutral state ($\tilde{g}uud$).  
 These charged gluebarinos are produced by cosmic
rays in the earth atmosphere \cite{VO}, and light gluino as LSP is 
excluded by the search for heavy hydrogen or by proton decay 
experiments (in case of quasistable gluino). In the case that the lightest 
gluebarino is neutral, see \cite{fa96}, the arguments of \cite{VO}
still work if a  neutral gluebarino forms a bound state with the nuclei.  
Thus, a light gluino is disfavored.

The situation is different if the gluino is heavy, 
$m_{\tilde{g}}\gsim 150~GeV$.  This gluino can be unstable
due to weak R-parity violation \cite{BJV} and have a lifetime 
$\tau_{\tilde{g}} \gsim 1$~yr, {\it i.e.\/}
long enough to be UHE carrier from remote parts of the Universe.  
Then the calculated relic density at the time 
of decay is not in conflict with the cascade nucleosynthesis and all 
cosmologically produced $\tilde{g}$-hadrons decayed up to the present time.
Moreover, the production of these gluinos by cosmic rays in the
atmosphere is ineffective because of their large mass.

\section{ Discussion and Conclusions}
We estimated the fluxes of UHE LSP produced by the decays of 
supermassive $X$-particles ($m_X > 10^{13}$~GeV). These particles can be  
Big Bang relics or can be produced by topological defects. 
We performed a simple Monte-Carlo simulation for the development of such 
a cascade. The fluxes of UHE LSP are numerically evaluated for two cases:
cosmic necklaces as example of topological defects, and supermassive 
Big Bang relic particles. 

Most attention is given to neutralino, $\chi$, as LSP. We calculated the   
$\chi N$-cross-sections at very high energies for two versions of 
"standard" MSSM with soft breaking terms: with universal scalar mass term
and non-universal one. 
The typical 
values of cross-sections at extremely high energies are 
$\sigma \sim 10^{-34}$~cm$^2$ 
and thus the detection of UHE neutralinos is a difficult task, which 
needs a special discussion. 
If the masses of the squarks are near their 
experimental bound \cite{pdg}, $M_{L,R} \sim 180$~GeV,
the cross-section can be 60 times higher.

What is important to realize is that neutralinos, which are probably
the most abundant form of matter in the Universe, 
can naturally exist in the form of high-energy particles. The signature
of UHE neutralino is a narrow resonance peak at energy 
$E\sim 10^9$~GeV$(M_e/1$~TeV$)^2$ (see Eq.(13)).

As another example of LSP we considered the gluino. This is a
hypothetical case, 
because in all elaborated SUSY models the gluino is not the LSP. Apart from 
a controversial mass window, $1- 4$~GeV, the gluino mass is limited from
below by accelerator experiments as 
$m_{\tilde{g}} \gsim 150$~GeV, while from above it is limited cosmologically 
($m_{\tilde{g}}\ll 10$~TeV).
The light gluino as LSP is further disfavored due to the upper limit on 
concentration of heavy hydrogen  or by searches for proton decay
\cite{VO}. The heavy unstable gluino with mass $m_{\tilde{g}}$ in the 
interval $150 - 1000$~GeV and the lifetime $\tau_{\tilde{g}} \gsim 1$~yr
is free from the above constraints and can be the carrier of UHE signal from
remote parts of the Universe. Heavy $\tilde{g}$-hadrons 
 behave in the earth atmosphere like 
penetrating particles: they have large cross-section, $\sigma \sim 1$~mb,
for very small energy transfers but very small cross-sections for large 
fractions of energy transfer $y \geq 0.1$. The light $\tilde{g}$-hadrons 
interact like the usual nucleons.

The energy spectrum of light $\tilde{g}$-hadrons
has the usual GZK-cutoff, while for  heavy ones the
energy of the GZK-cutoff is shifted to larger energies as shown in 
Fig.~2. 
In conclusion, we think that gluino is disfavored both as LSP and 
the carrier of UHE signal.

\begin{ack}
We acknowledge with gratitude the participation of A. Vilenkin in the 
initial stage of this work.
We are grateful to V. De Alfaro, A. Kaidalov and V. Matveev for useful 
discussions and to M. Nagano for providing us with the experimental data
used in Fig.~1 and 2 and for helpful remarks.
M.K. was supported by a Feodor-Lynen scholarship of the Alexander
von Humboldt-Stiftung.

{\em Note added:}
At the moment of the submission of this paper, a paper by J. Adams
{\it et al.\/} (hep-ex/9709028) has appeared which excludes a light
glueballino in the mass and life time ranges of $1.2-4.6$~GeV and
$2\cdot 10^{-10} - 7\cdot 10^{-4}$~s. While these data are not
consistent with the model of Farrar {\it et al.\/} \cite{fa96,ch97},
they do not exclude a long-lived gluino with $\tau\gsim 1$~yr which we
critically discussed.  
\end{ack}


\begin{thebibliography}{99}


\bibitem{fa96}
G. R. Farrar, Phys. Rev. Lett. {\bf 76} (1996) 4111 and references
therein cited.

\bibitem{pdg}
Particle Data Group, Phys. Rev. {\bf D54} (1996) 1.

\bibitem{aleph}
Aleph collaboration, CERN-PPE-97/002, to be published in Z. Phys. C;
G. R. Farrar, hep-ph/9707467.

\bibitem{ch97}
D. J. H. Chung, G. R. Farrar and E. W. Kolb, astro-ph/9707036,
see also Ref. \cite{fa96}.

\bibitem{mo97}
R. N. Mohapatra and S. Nussinov, hep-ph/9708497. 

\bibitem{aur}
G. Auriemma, L. Maiani and S. Petrarca, 
Phys. Lett. {\bf B164} (1985) 179.
\bibitem{BI}
V. S. Berezinskii and B. L. Ioffe, 
Sov. Phys. JETP {\bf 63} (1986) 920.

\bibitem{turn} 
M. S. Turner, Nature, 297 (1982) 379; 
J. H. MacGibbon and B. J. Carr, Astrophys. J. 371 (1991) 447.
   
\bibitem{nov}
I. D. Novikov et al., Astron. Astrophys. 80, 104 (1979);
B. J. Carr, J. H. Gilbert and J. E. Lidsey, Phys.Rev. D50, 4853 (1994).


\bibitem{td}
A. Vilenkin and E. P. S. Shellard,
{\em Cosmic Strings and other Topological Defects\/},
Cambrige University Press, 1994;
M. B. Hindmarsh and T. W. B. Kibble, Rep. Prog. Phys. {\bf 58} (1995) 477.

\bibitem{CR_TD}
E. Witten, Nucl. Phys. {\bf B247} (1985) 557;
C. T. Hill, D. N. Schramm and T. P. Walker, Phys. Rev. {\bf D36} (1987) 1007;
P. Bhattacharjee and G. Sigl, Phys. Rev. {\bf D51} (1995) 4079.

\bibitem{BV}
V. Berezinsky and A. Vilenkin, astro-ph/9704257.

\bibitem{BKV}
V. Berezinsky, M. Kachelrie{\ss} and A. Vilenkin, astro-ph/9708217;
see also
V. A. Kuzmin and V. A. Rubakov, preprint astro-ph/9709187.

\bibitem{casc}
V. Berezinsky, M. Kachelrie{\ss} and R. Sang, in preparation.

\bibitem{photino}
V. S. Berezinskii, E. V. Bugaev and E. S. Zaslavskaya, 
Nucl. Phys. {\bf B272}, 193 (1986).

\bibitem{data}
M. Nagano, private communication.

\bibitem{be96}
V. Berezinsky {\it et al.}, Astropart. Phys. 5 (1996) 1.

\bibitem{VO}
M. B. Voloshin and L. B. Okun, Sov. J. Nucl. Phys. 43 (1986) 495.

\bibitem{anom}
T. K. Hemmick {\it et al.}, Phys. Rev. {\bf D 41} (1990) 2074.

\bibitem{BJV}
V. Berezinsky, A. S. Joshipura and J. W. F. Valle, hep-ph/ 9608307, to be 
published in Phys. Rev. D.

\end{thebibliography}
\end{document}